\documentclass{aa}
\usepackage{txfonts}
\usepackage{graphicx}
\def\msun{\hbox{M$_\odot$}}

\begin{document}

\title{Astrophysical properties of newly discovered Magellanic Cloud star clusters}

\author{Andr\'es E. Piatti\inst{1,2}\thanks{\email{andres.piatti@unc.edu.ar}}}

\institute{Instituto Interdisciplinario de Ciencias B\'asicas (ICB), CONICET-UNCUYO, Padre J. Contreras 1300, M5502JMA, Mendoza, Argentina;
\and Consejo Nacional de Investigaciones Cient\'{\i}ficas y T\'ecnicas (CONICET), Godoy Cruz 2290, C1425FQB,  Buenos Aires, Argentina\\
}

\date{Received / Accepted}

\abstract{New star cluster candidates projected toward  the Large and Small
Magellanic Clouds (L/SMC) have been 
recently discovered from relatively deep imaging surveys. We here conduct a sound analysis 
of 24 star cluster candidates located in the outer regions of the L/SMC using PSF photometry 
produced by the Survey of the Magellanic  Stellar History. With only one exception, the studied 
objects resulted to be
genuine stellar aggregates. We conclude on their physical reality once their observed
color-magnitude diagrams (CMDs) were statistically decontaminated by the presence of field 
stars,  and the resulting cleaned CMDs for stars with assigned membership probabilities
higher than 50\% were compared with synthetic CMDs generated for thousand combinations of 
ages, distances, metallicities, star cluster mass and binary fractions. The parameter of the 
best-matched synthetic CMDs obtained from a likelihood approach were adopted as the star cluster astrophysical  properties. The present star cluster sample spans a wide range of distances, from
those star clusters located in front of the LMC, to those along the onset of the Magellanic Bridge, 
up to those behind the SMC. Their ages reveal different formation episodes that took place along 
the galaxy formation and others as a consequence of the galaxy interactions. From their estimated
metallicities and ages, we speculate with the possibility that relatively metal-deficient gaseous
flows have existed between these galaxies during nearly the last one Gyr
 (log(age /yr) $\approx$ 9.0), that made possible
the formation of young star clusters in the galaxy peripheries. Despide the L/SMC interactions,
the studied star clusters are similar or more massive than their counterparts in the Milky Way,
which suggests that tidal effects are relatively more important in our Galaxy.}
 

\titlerunning{}

\authorrunning{Andr\' es E. Piatti }

\maketitle

\markboth{Andr\' es E. Piatti: }{Magellanic Clouds star clusters}

\section{Introduction}

Recent imaging surveys of the Magellanic Clouds have allowed the community to embark in searches 
of relatively compact, small and poorly populated star clusters \citep[see, Table 1 in][]{maiaetal2019}. 
Imaging visual inspections \citep[e.g.,][]{bicaetal2020} or machine learning techniques 
\citep[e.g.,][]{cernyetal2020} revealed the existence of stellar overdensities proposed as star 
clusters spread throughout the surveyed areas. Several of these star cluster candidates have not been 
confirmed as genuine physical systems \citep[][]{petal14b,p18c}, while others turned out to be 
star clusters with properties that have defied our previous knowledge about the Magellanic Clouds 
formation, structure, dynamics and chemical evolution \citep[][]{petal2016,piattietal2018a,gattoetal2020}.

By using wide-field high-quality images released in advance from the  Survey of the Magellanic Stellar 
History (SMASH) \citep{nideveretal2017a}, \citet{p17a} used density kernel estimators with
physically meaningful bandwidths  to detect the smallest and/or less dense star clusters in the
Magellanic Clouds. He found out 24 new star cluster candidates (see his Table 1), most of them located 
in the outer regions of both Clouds, thus reinforcing previous suggestions that those regions were less
explored in the past. He did not estimated their fundamental parameters, which means that the reality of
these candidates as stellar aggregates needs to be confirmed. The SMASH DR2 is now
publicly available from the portal of the Astro Data Lab\footnote{https://datalab.noao.edu/smash/smash.php},
which is part of the Community Science and Data Center of NSF’s National Optical Infrared Astronomy 
Research Laboratory. Therefore, we have now the chance of studying in details these candidates,
providing for the first time with their astrophysical properties.
SMASH  is a community Dark Energy Camera (DECam)  survey of the Magellanic
Clouds mapping 480 deg$^2$ (distributed over $\sim$2400 deg$^2$ at $\sim$20$\%$
 filling factor, which complements the 5000 deg$^2$ Dark Energy Survey’s partial coverage of the Magellanic periphery) to $\sim$24th mag $griz$ (and $u$ $\sim$23),
with the goal of identifying broadly distributed, low surface brightness stellar 
populations associated with the stellar halos and tidal debris of the Magellanic Clouds.
SMASH will also derive spatially-resolved star formation histories covering all ages 
out to large radii of the Magellanic Clouds that will further complement our 
understanding of their formation. DECam is a wide-field optical imager 
(FOV = 2.2$\times$2.2 deg$^2$, scale = 0.263 arcsec/pix )attached at 
Cerro Tololo  Interamerican Observatory Blanco 4m telescope.

Beyond the usefulness of enlarging the sample of well-studied Magellanic Cloud star clusters, the fact
that \citet{p17a}'s new candidates are located in the outer regions of both Clouds is of particular
interest. The outer regions of the Large and Small Magellanic Clouds (L/SMC) have been primary
scenarios of the interaction between both galaxies, namely: star clusters were stripped off
\citep{carpinteroetal2013}; new stars clusters formed \citep{piattietal2018b}; and ancient star clusters 
were thought to be found \citep{piattietal2019}. Hence, star clusters located in these regions challenge 
our ability of disentangling the early star cluster formation episodes from more recent formation events. 
Such a distinction is necessary for a better understanding of the age-metallicity relationships
and metallicity gradients observed in these galaxies \citep{pg13}.

In this work, we analyze SMASH data sets for the 24 star cluster candidates identified by \citet{p17a}.
Section 2 deals with the decontamination of the color-magnitude diagrams (CMDs) from field star
contamination and the estimate of ages, distances and metallicities. The implications of the resulting
parameters in the context of the formation and interaction of both Magellanic Clouds are discussed
in Section 3. Finally, Section 4 summarizes the main conclusions of this work.

\section{Fundamental parameters of star clusters}
\subsection{Cleaning color-magnitude diagrams}

We retrieved R.A and Dec. coordinates, PSF $g,i$ magnitudes and their respective errors, $E(B-V)$
interstellar reddening and $\chi$ and {\sc sharpness} parameters of stellar sources distributed within
circles with radii of 6$\arcmin$ (the star cluster candidates are smaller than $\sim$ 0.3$\arcmin$)
from the Astro Data Lab. In order to assure the selection of point sources, we applied the following
filters: 0.2 $\le$ {\sc sharpness} $\le$ 1.0  and $\chi$ $<$ 0.5, so cosmic rays, 
bad pixels, galaxies,  and unrecognized double  stars were excluded. 
{\sc sharpness} and $\chi$ are image quality diagnostic parameters used by 
{\sc daophot}.

\begin{figure}
\includegraphics[width=\columnwidth]{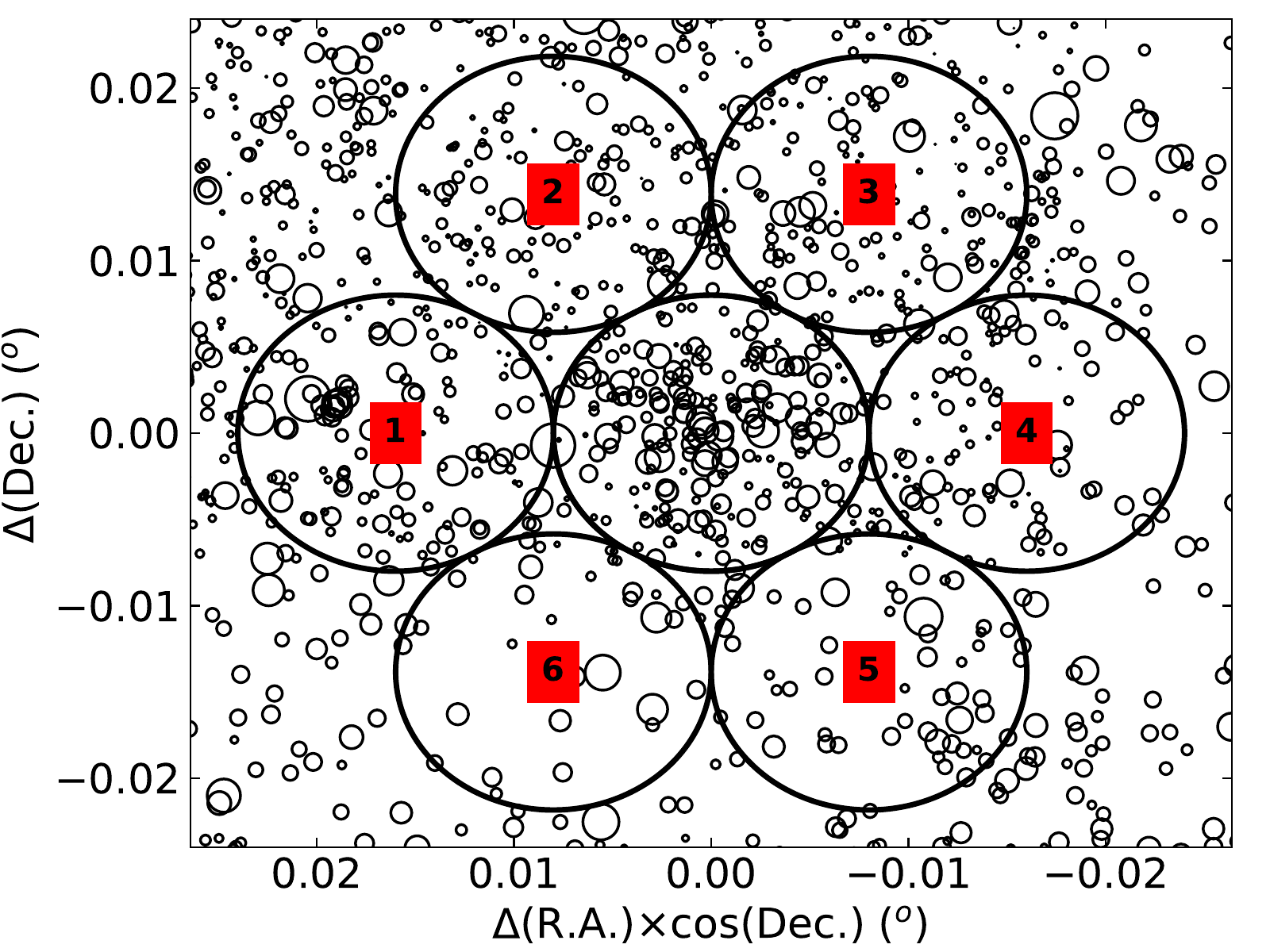}
\caption{Schematic chart centered on Field\,16-02. The size of the symbols is proportional to the $g$ brightness. The radius of the superimposed circles equals the adopted cluster’s radius
(see Table~\ref{tab:tab1}). Six labeled reference star fields distributed around the star cluster circle are also drawn.
}
\label{fig:fig1}
\end{figure}

We carefully monitor the  contamination of field stars in the star cluster CMDs by using
6 different star field CMDs built from stars distributed in circular areas of equal size as the cluster 
area distributed around it (see, Fig.~\ref{fig:fig1}). This is because the star field varies in stellar density as well as in the 
distribution of brightness and color of its stars from one place to the other.  The chosen regions are 
meant to include any possible star field population and reddening variation around the star clusters.
Because the star cluster candidates are relatively small and would seem to contain a relative
small number of stars \citep[see, Figure 3 in][]{p17a}, we decided to clean cluster areas with
radii slightly larger than the readily visible clusters' dimensions. Thus, we minimize the presence
of potential residuals from the cleaning procedure when building the cleaned star cluster CMDs.

Field star contamination plays an important role when analyzing Magellanic Cloud star cluster
CMDs. Because of the galactocentric distances of both galaxies, star cluster and fiel star
sequences in the CMDs can be superimposed. This means that it is not straightforward 
to consider a star a cluster member from its lone position in the CMD. Such an ambiguity 
can be solved, sometimes, with additional  information of proper motions, radial velocities, and/or 
chemical abundances of individual stars  Unfortunately, in  the case of our  star cluster candidates, 
{\it Gaia} DR2 proper motions  \citep{gaiaetal2016,gaiaetal2018b} are unavailable for several stars
concentrated in very small regions or they are unreliable because of their apparent low brightness. 
For this reason, we exploit the photometry of reference field stars to decontaminate the
star cluster CMDs.

In general terms, the reference star field is placed adjacent to the star cluster field, but not 
too  far from it, so that it can be a suitable representative of the star field projected along the
line-of-sight (LOS) of the star cluster. Frequently, the assumption of homogeneity in the stellar density and
in the distribution of luminosities and effective temperatures of field stars across the star 
cluster  field and around it is adopted. This means that field stars located along the LOS of the 
star cluster can be mimic in number and astrophysical properties by those located along a 
direction slightly shifted from that toward the star cluster. However, even though the
star cluster is not projected onto a crowded star field or is not affected by differential
reddening, it is highly possible to find differences throughout the star cluster surrounding
field. Bearing in mind the above considerations, we decided to clean the star field 
contamination in the star cluster CMDs by using, at a time, the 6 different devised reference 
star field areas described above. 

We follow three main steps while decontaminating the star cluster CMDs. On the one
hand, we properly represent each reference star field by considering simultaneously its stellar 
density and the observed distribution of its stars in luminosity and effective temperature
(number of stars per CMD mag and color units). Then, we statistically subtract the reference star 
field from the star cluster CMD and, finally, we assign membership probabilities from the
consideration of the six different resulting  cleaned star cluster CMDs (one cleaned CMD per 
reference star field CMD). Stars with relatively high membership probabilities that are located
along a single theoretical isochrone (corresponding to an age, distance and metallicity) are
considered cluster members. The method was devised by \citet{pb12} and successfully
used elsewhere to decontaminate CMDs of star clusters projected on to crowded fields in
the Milky Way and the Magellanic Clouds \citep[see, e.g.][and references therein]{petal2016,p2018,piattietal2020}.

\begin{figure}
\includegraphics[width=\columnwidth]{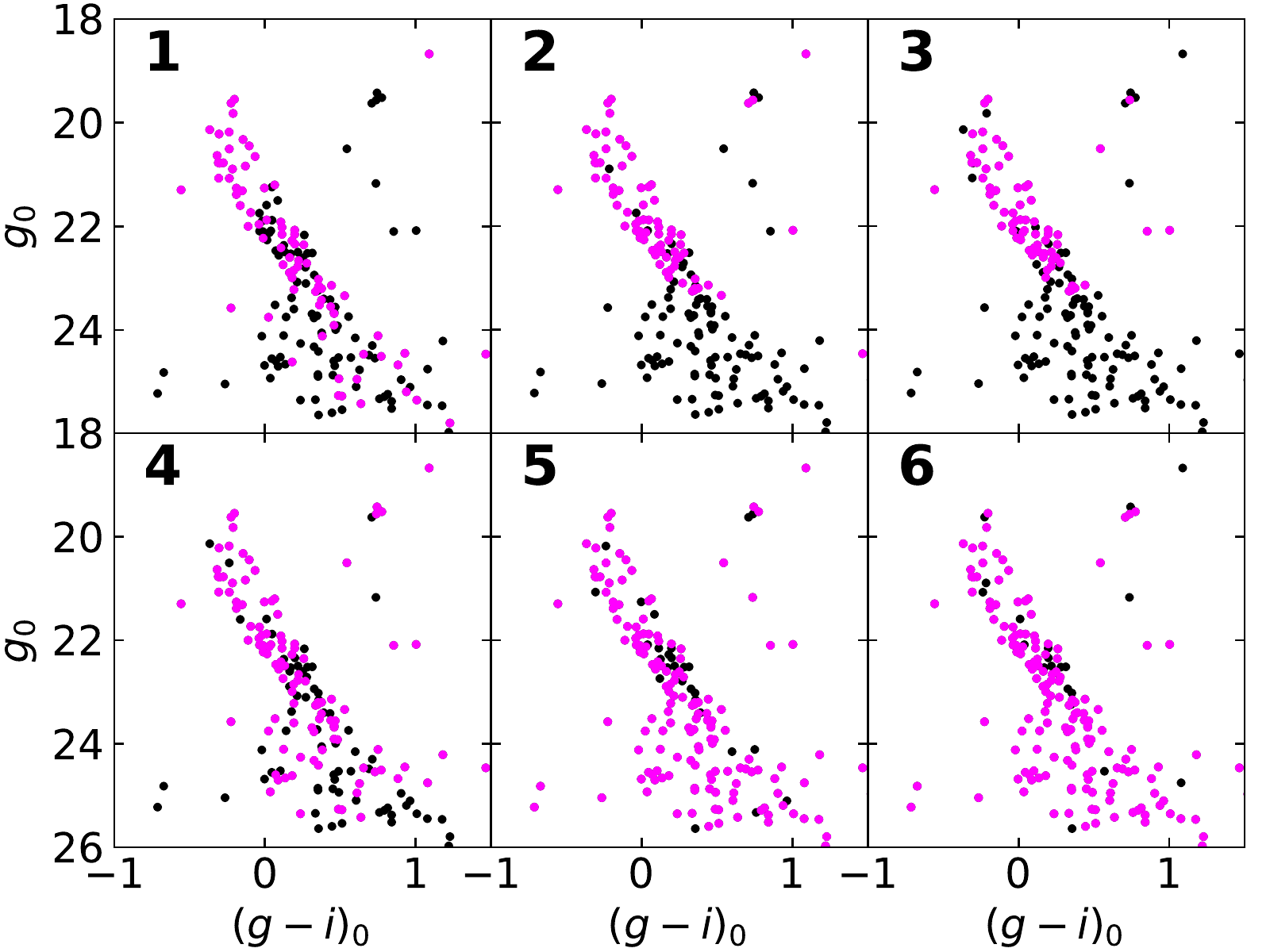}
\caption{Color-magnitude diagram of Field\,16-02. Black points represent all the measured stars located within the cluster radius. Magenta points represent the stars that remained unsubtracted after the CMD cleaning procedure. The reference star field used to decontaminate the star cluster CMD is indicated at the top-left margin (see also Fig.~\ref{fig:fig1}).}
\label{fig:fig2}
\end{figure}

\begin{figure}
\includegraphics[width=\columnwidth]{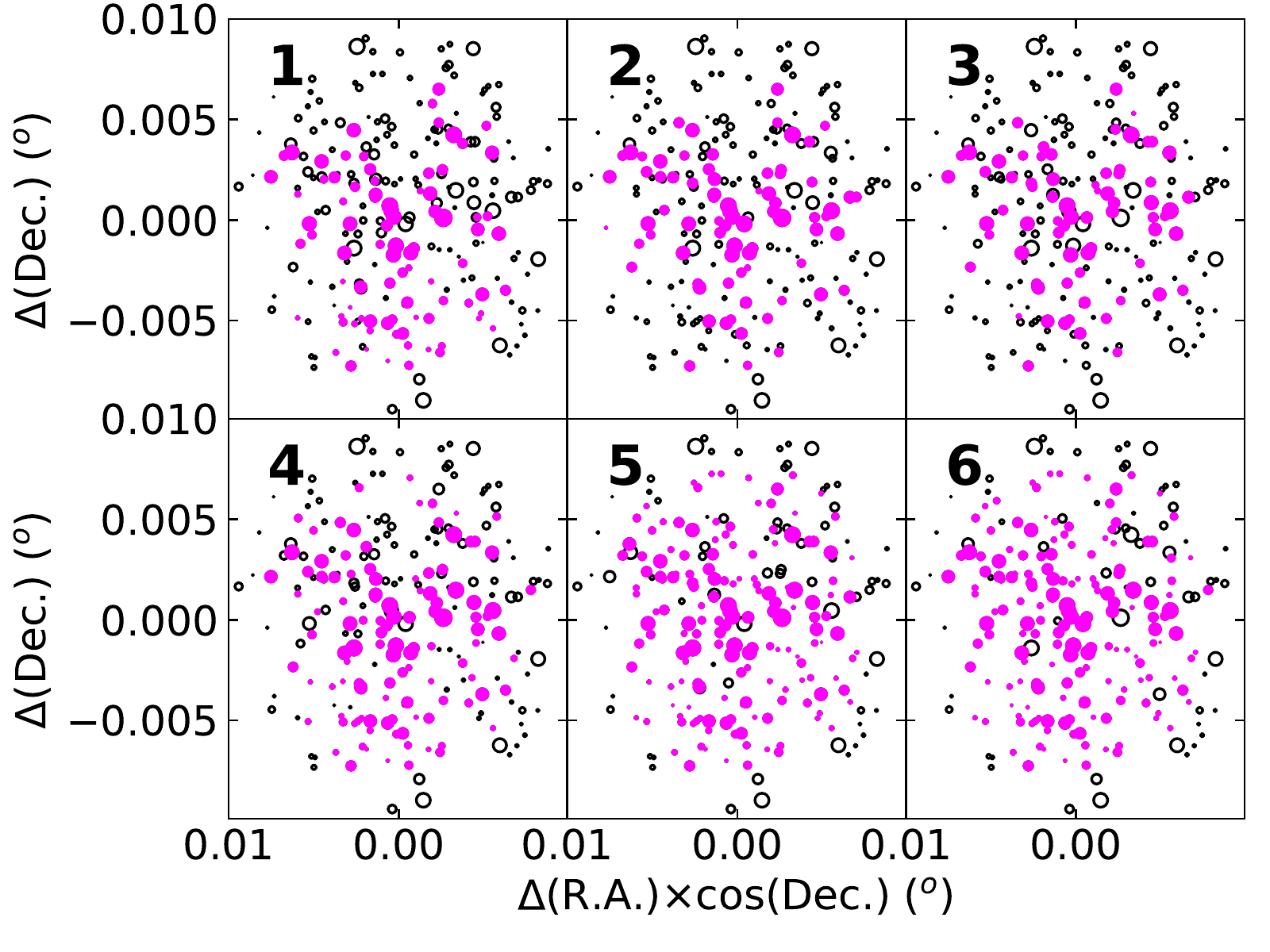}
\caption{ Chart of the stars in the field of Field\,16-02. The size of the symbols is proportional to the $g$ brightness of the star. Open black circles represent all the measured stars located in the cluster circle. Filled magenta circles represent the stars that remained unsubtracted after the CMD cleaning procedure. The reference star field used to decontaminate the star cluster CMD is indicated at the top-left margin (see also Fig.~\ref{fig:fig1}).}
\label{fig:fig3}
\end{figure}

We subtract from the star cluster CMD a number of stars equal to that in the reference star field.
If we subtracted less or more stars, we could conclude on the presence of a more populous object
or the absence of a real aggregate, respectively. The distribution of magnitudes  and colors of the 
subtracted stars needs in addition to resemble that of the reference star field. With the aim of 
avoiding stochastic effects caused by very few field stars distributed in less populated CMD regions, 
appropriate ranges of magnitudes and colors around the CMD positions of field stars are 
advisable to be used. Thus, it is highly probable to find a star in the star cluster CMD
with a magnitude and a color within those boundaries around the magnitude and color
of each field star. In the case that more than one star is located inside that delimited CMD
region, the closest one to the center of that (magnitude, color) box is subtracted. We started here
with boxes of ($\Delta$$g$, $\Delta$$(g-i)$) = (2.0 mag,1.0 mag) centered on the magnitude and color
values of each reference field star. We based our analysis on dereddened
CMDs, so we first corrected by interstellar extinction the $g$ and $i$ magnitudes using the
$E(B-V)$ values provided by SMASH and the $A_\lambda$/$E(B-V)$ ratios, for $\lambda$ =
$g,i$, given by  \citet{abottetal2018}. The photometric errors are also taken into account while 
searching for a star to be
subtracted from the star cluster CMD. With that purpose, we iterate up to1000 times 
the comparison between the magnitude and color of the reference field star and
those of the stars in the star cluster CMD. If a star in the star cluster CMD falls
inside the box defined for the reference field star, we subtract that star. The iterations
are carried out by allowing the magnitude and color of the star in the star cluster CMD 
takes smaller or larger values than the mean ones according to their respective
errors. Figure~\ref{fig:fig2} illustrates the results of the decontamination of field stars using the six different reference star fields depicted in Fig.~\ref{fig:fig1}.  The spatial distribution of these stars is shown in Fig.~\ref{fig:fig3}.

\begin{figure*}
\includegraphics[width=\textwidth]{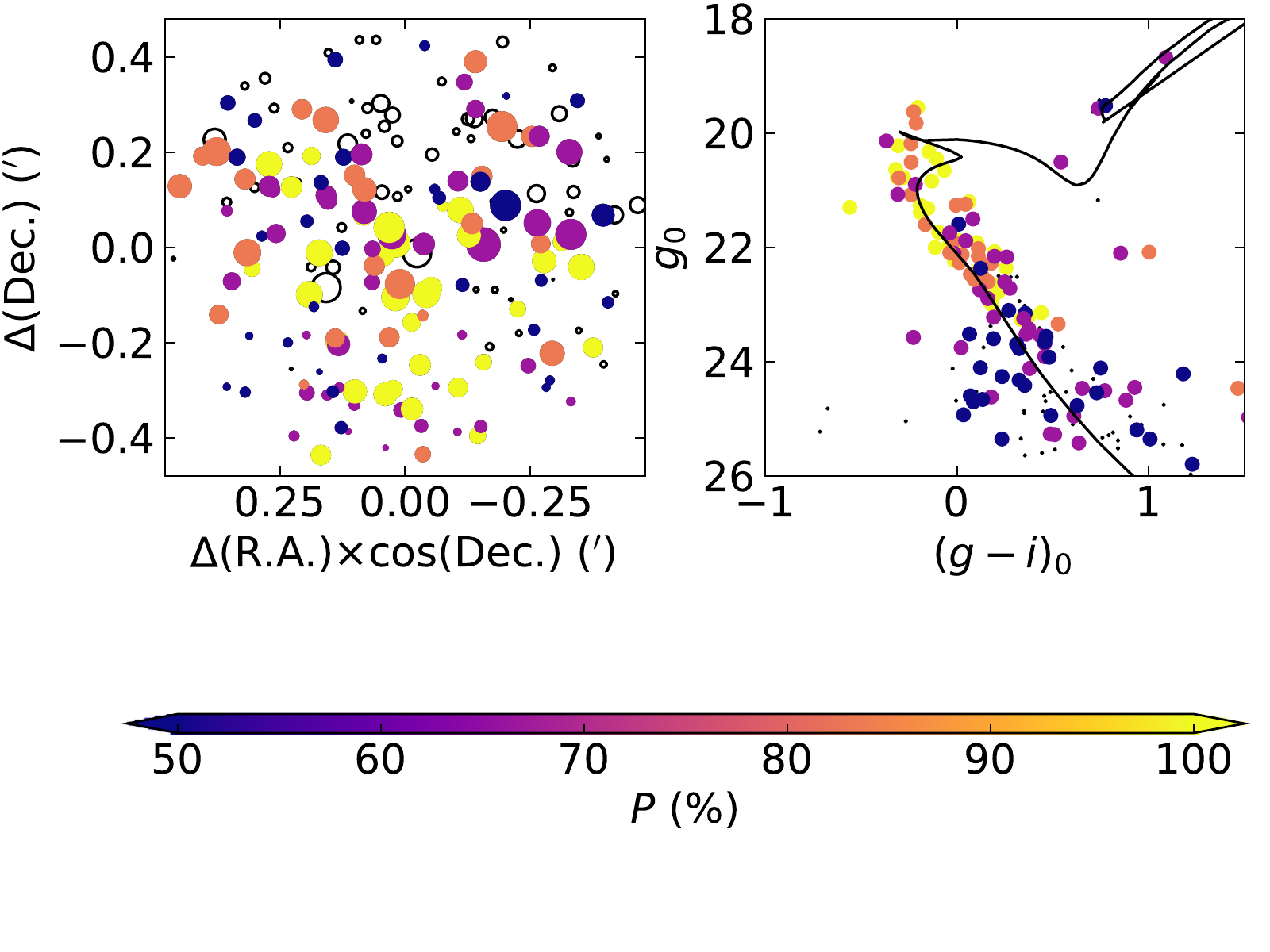}
\caption{Spatial distribution (left panel) and corresponding intrinsic CMD (right panel) of stars 
measured by SMASH in the field of Field\,16-02. Symbols for stars with membership
probabilities $P \ge$ 50$\%$  are painted according to the color bar. The best fitted theoretical
isochrone is also superimposed in the cluster CMD.}
\label{fig:fig4}
\end{figure*}

We finally assign a membership probability to each star that remain unsubtracted
after the decontamination of the star cluster CMD. Because unsubtracted stars
vary from one cleaned CMD to the other, we define the membership probability
$P$ ($\%$) = 100$\times$$N$/6, where $N$ represent the number of times a
star was not subtracted during the 6 different CMD cleaning executions. 
Figure~\ref{fig:fig4} illustrates the spatial distribution and corresponding CMD
of stars measured in the field of the star cluster Field\,16-02, painted according
to their membership probabilities $P$. We applied the above
cleaning procedure to the remaining 23 star cluster candidates discovered by 
\citet{p17a}. The resulting cleaned star cluster CMDs and the respective spatial distributions 
of the measured stars are included in the Appendix. We found Field\,30-02 not 
to be a real star cluster, but a chance grouping of stars. 

\subsection{Estimating star cluster fundamental parameters}

At first glance, the cleaned star cluster CMDs reveal objects seemingly spanning  young to moderately 
old ages. Because the age estimate depends on the star cluster metallicity and distance, we employ
routines of the Automated Stellar Cluster Analysis code \citep[\texttt{ASteCA,}][]{pvp15}
to derive all of them simultaneously. \texttt{ASteCA} is a suit of tools designed to analyze 
data sets of star clusters in order to determine their basic parameters. We thus 
obtain a synthetic CMD that best matches the star cluster CMD. The metallicity, age, distance, star 
cluster present mass and binary fraction associated to that generated synthetic CMD are adoped
as the best-fitted star cluster parameters. 

We start by using the theoretical isochrones computed by \citet{betal12} for the SMASH photometric
system. We downloaded theoretical isochrones for different metallicities values, from $Z$ = 0.000152
([Fe/H]=-2.0 dex) up to 0.030152 ([Fe/H]=0.30 dex) in steps of $\Delta Z$=0.001. This metallicity
range cover almost all the metallicity regime of the Magellanic Clouds \citep{pg13}. This is an
important consideration, because the studied star clusters lie in the outer regions of the L/SMC,
where  metal-poor old and metal-rich young objects formed at the galaxy formation and 
galaxy interaction, respectively. As for ages,
we downloaded isochrones from log(age /yr)=6.0 (1 Myr) up to 10.1 (12.5 Gyr) in steps of
$\Delta$log(age /yr)=0.05. In total, we gathered nearly 2500 different theoretical isochrones.

The input data sets consist in intrinsic magnitudes $g_0$ and dereddened colors $(g-i)_0$ for
all the stars with membership probabilities $P \ge$ 50$\%$, i.e., all colored points in Figures~
\ref{fig:fig4} and \ref{fig:figa1}-\ref{fig:figa8}, with their respective uncertainties. For generating
the synthetic CMDs, we adopted the initial mass function of \citet{kroupa02}; a minimum mass
ratio for the generation of binaries of 0.5; and a range of true distance moduli from 18.0 mag
(40 kpc) up to 19.5 mag (80 kpc). We explore the parameter space of the synthetic CMDs 
through the minimization of the likelihood function defined by \citet{tremmeletal2013} using
a parallel tempering Bayesian MCMC algorithm. Errors in the obtained parameters are
estimated from the standard bootstrap method described in \citet{efron1982}. We refer the reader
to the work of \citet{pvp15} for details concerning the implementation of these algorithms.
Table~\ref{tab:tab1} lists the resulting parameters for the entire star cluster sample. We illustrate
the performance of the parameter matching procedure by superimposing the isochrone corresponding to the
best-matched synthetic CMD to the cleaned star cluster CMDs (see,  Figures~\ref{fig:fig4} and 
\ref{fig:figa1}-\ref{fig:figa8}).

\begin{table*}
\begin{center}
\caption{Fundamental parameters of Magellanic Clouds clusters: radius of the cleaned star cluster area ($r$); true distance modulus
($(m-M)_0$); age; metallicity ([Fe/H]); star cluster mass; and binary fraction ($q$).}
\label{tab:tab1}
\begin{tabular}{@{}lcccccccccccc}\hline\hline
Star cluster$^a$ & R.A. & Dec. & $r$          & $(m-M)_0$& log(age /yr) & [Fe/H] & Mass         &  $q$  \\
                          & (deg) & (deg)&  (arcmin) &  (mag)       &                   & (dex)   & ($\msun$) &        \\\hline
Field\,4-01 &8.254   & -72.989  & 0.30 &18.73$\pm$0.22 & 9.47$\pm$0.16 &	-0.97$\pm$0.31  &806$\pm$626&	0.44$\pm$0.27 \\
Field\,10-01 &16.895 &-72.164  & 0.30 &18.76$\pm$0.24 & 7.54$\pm$0.49 &	-0.92$\pm$0.05  &	338$\pm$297&	0.39$\pm$0.25 \\ 
Field\,10-02 &18.098 &-72.292 &0.30  & 18.92$\pm$0.29 & 8.27$\pm$0.21 &	-0.53$\pm$0.27 &	177$\pm$38&	0.51$\pm$0.28  \\
Field\,10-03 &16.194 &-72.716 &0.30  & 18.87$\pm$0.36 & 8.11$\pm$0.18 &	-0.43$\pm$0.26 &	358$\pm$147&	0.51$\pm$0.25  \\
Field\,11-01 &15.982 &-72.933 &0.30  & 19.04$\pm$0.37 & 7.87$\pm$0.18 &	-0.52$\pm$0.34 &	926$\pm$586&	0.53$\pm$0.28	 \\
Field\,11-02 &15.325 &-73.465 &0.30  & 18.56$\pm$0.34 & 8.15$\pm$0.13 &	-0.44$\pm$0.25 &	628$\pm$250&	0.53$\pm$0.24  \\
Field\,11-03 &16.048 &-74.299 &0.20 & 19.10$\pm$0.26 &  8.96$\pm$0.62 &   -0.71$\pm$0.45 &	534$\pm$341&	0.52$\pm$0.29 \\ 
Field\,12-01 &18.427 &-74.753 &0.30  & 18.42$\pm$0.24 & 9.74$\pm$0.15 &	-0.98$\pm$0.44	&	155$\pm$44&	0.54$\pm$0.29  \\
Field\,15-01 &20.753 &-73.218  & 0.25 & 18.72$\pm$0.21 & 8.05$\pm$0.45 &	-0.80$\pm$0.43 &	208$\pm$85&	0.37$\pm$0.25 \\ 
Field\,16-01 &22.471 &-74.846  & 0.50 & 19.01$\pm$0.15 & 9.09$\pm$0.07 &	-0.53$\pm$0.13  &	390$\pm$58&	0.53$\pm$0.25  \\
Field\,16-02 &22.461 &-74.682 &0.50  &18.87$\pm$0.31 & 9.19$\pm$0.13 &	-0.83$\pm$0.30& 	1085$\pm$198&0.68	$\pm$0.19 \\ 
Field\,30-01 &72.066 &-69.672 &  0.30 & 18.58$\pm$0.26 & 7.88$\pm$0.39 &	-0.44$\pm$0.27 &	238$\pm$102&	0.51$\pm$0.28  \\
Field\,40-01 &80.275 &-71.913 & 0.30 &18.64$\pm$0.41 & 7.34$\pm$0.41 &	-0.78$\pm$0.18  &	601$\pm$377&	0.58$\pm$0.30  \\
Field\,40-02 &81.364 &-72.639 & 0.30 & 18.42$\pm$0.25 & 8.76$\pm$0.25 &	-0.78$\pm$0.37  &    	186$\pm$60&	0.60$\pm$0.28  \\ 
Field\,40-03 &78.142 &-72.626  &  0.40 & 18.52$\pm$0.28 & 8.89$\pm$0.27 &	-0.49$\pm$0.28 & 	116$\pm$16&	0.58$\pm$0.28 \\ 
Field\,40-04 &79.484 &-73.557  &0.30 & 18.62$\pm$0.40 & 9.01$\pm$0.23 &	-0.26$\pm$0.17  & 	213$\pm$84&	0.54$\pm$0.29  \\
Field\,40-05 &80.489 &-73.513 &0.50 & 18.40$\pm$0.32 & 8.96$\pm$0.16 &	-0.73$\pm$0.17& 	215$\pm$100&	0.63$\pm$0.26  \\
Field\,40-06 &78.088 &-73.270 &0.30 & 18.83$\pm$0.41 & 8.66$\pm$0.17 &	-0.37$\pm$0.23& 	254$\pm$118&	0.50$\pm$0.29  \\
Field\,40-07 &81.081 &-73.091 &0.30 & 18.69$\pm$0.33 & 8.55$\pm$0.72 &	-0.26$\pm$0.15 & 	342$\pm$372&	0.56$\pm$0.28	 \\ 
Field\,44-01 &82.674 &-75.668 &0.30 & 18.50$\pm$0.37 & 9.34$\pm$0.25 &	-0.38$\pm$0.24& 	389$\pm$340&	0.53$\pm$0.29  \\
Field\,44-02 &81.523 &-75.416 &0.25 & 18.42$\pm$0.29 & 9.35$\pm$0.21 &	-0.54$\pm$0.22&	217$\pm$196&	0.53$\pm$0.28  \\
Field\,51-01 &87.308 &-70.6737 &0.30 & 18.57$\pm$0.26 & 7.87$\pm$0.38 &	-0.78$\pm$0.47& 	262$\pm$122&	0.55$\pm$0.29  \\
Field\,55-01 &96.733 &-70.3008  &0.20 &	18.56$\pm$0.34 &   8.70$\pm$0.29 & -0.48$\pm$0.18&  	154$\pm$46&	0.52$\pm$0.28 \\\hline
\end{tabular}
\end{center}
$^a$ Star cluster identifications are from \citet[][Table\,1]{p17a}.

\end{table*}

\section{Analysis and discussion}

The advantage of playing with thousands of synthetic CMDs allows a larger number of free
parameters to be fitted. This is the case of the true distance modulus. In dealing with
theoretical isochrones fitted to star cluster CMDs, a mean distance modulus is frequently
adopted, because the combination of the Magellanic Cloud distances with their 
respective LOS disk depths implies a variation of the distance moduli -bearing in mind that any 
star cluster could be placed in front of or behind the L/SMC-  of  $\Delta (m-M)_0 \sim$ 0.2 mag. 
The latter comes from considering for the LMC: $(m-M)_0 = 18.49 \pm 0.09$ mag \citep{dgetal14} 
and $<$LOS$>$ = 3.44$\pm$1.16 kpc \citep{ss09}, and for the SMC: $(m-M)_0 = 18.98 \pm 0.03$ 
mag \citep{graczyketal2020} and $<$LOS$>$ = 6.0$\pm$1.7 kpc \citep{cetal01}. This difference is much
smaller than the difference in absolute magnitude between two closely spaced isochrones with 
$\Delta$(log(age /yr)= = 0.1 (a typical age error), so that adoption of a unique value for the 
distance modulus does not dominate the final error budget incurred in matching
isochrones to the star cluster CMDs. However, the Magellanic Clouds are more extended
than previously estimated, showing tidally-induced warps, substructures and tidal distortions in
the peripheries, etc \citep{mackeyetal2016,mackeyetal2017,choietal2018,mackeyetal2017}.
Therefore, the use of the true distance modulus as a free parameter in the likelihood approach 
helps us to place the studied star clusters more accurately.  

\begin{figure}
\includegraphics[width=\columnwidth]{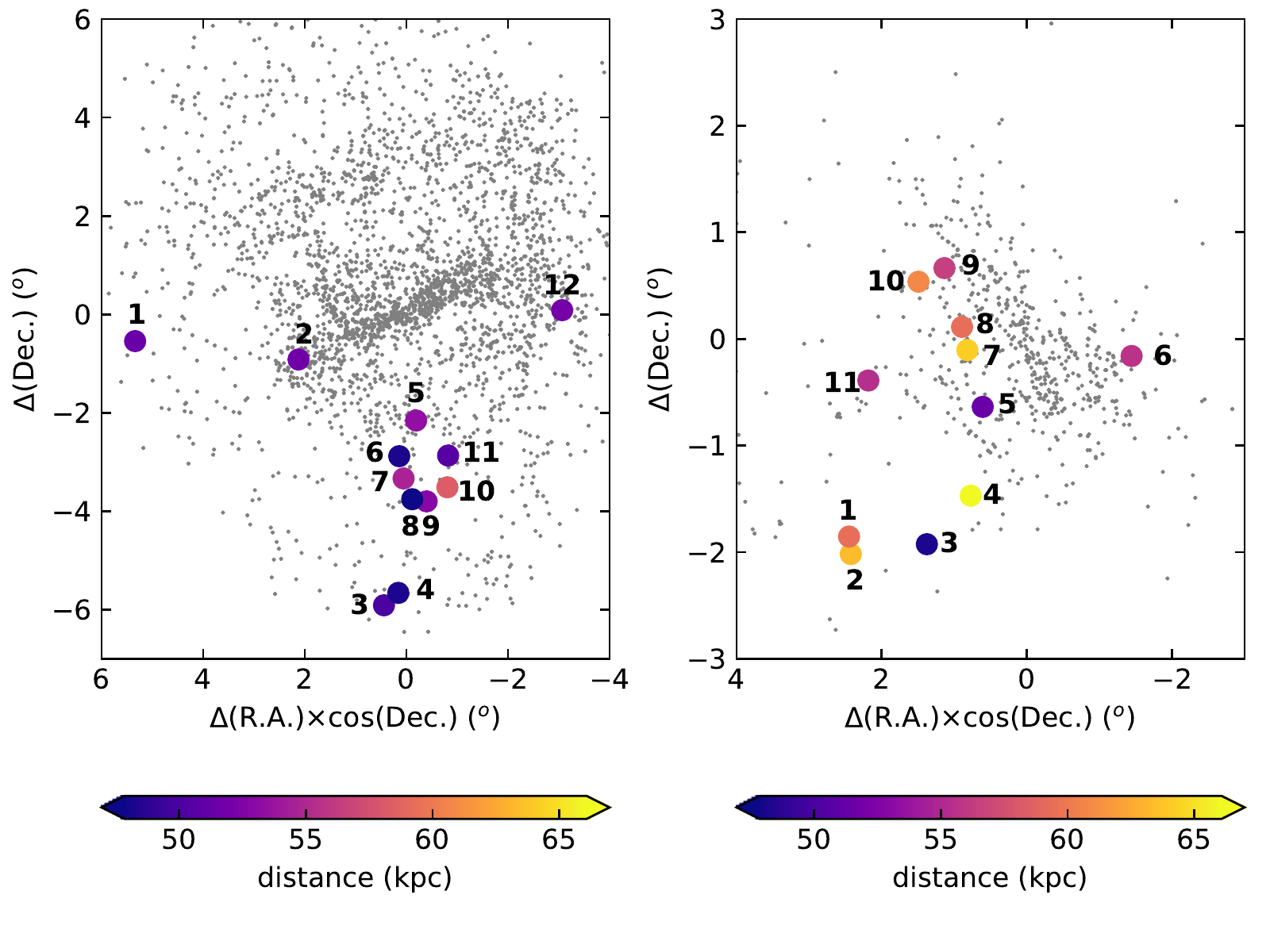}
\caption{ Spatial distribution of the Magellanic Clouds clusters. North is up and East to the
left. Gray points are star clusters in the \citet{betal08}'s catalog, while color-coded large circles represent the studied star cluster sample. {\it Left LMC:} 1= Field\,55-01; 2= Field\,51-01; 3= Field\,44-01; 
4= Field\,44-02; 5= Field\,40-01; 6= Field\,40-02; 7= Field\,40-06; 8= Field\,40-04; 9= Field\,40-05;
10= Field\,40-07; 11= Field\,40-03; 12= Field\,30-01. {\it Right SMC:} 1= Field\,16-02; 2= Field\,16-01;
3= Field\,12-01; 4= Field\,11-03; 5= Field\,11-02; 6= Field\,4-01; 7= Field\,11-01; 8= Field\,10-03; 9= Field\,10-01; 10= Field\,10-02; 11= Field\,15-01.}
\label{fig:fig5}
\end{figure}

The resulting spatial distribution of the studied star clusters is depicted in Figure~\ref{fig:fig5}.
For comparison purposes, we included as reference the positions of star clusters cataloged
by \citet{betal08}. They clearly delineate the bars, arms, outer disks, bridge, etc. As can be seen,
the present star cluster sample consists of objects spread out across the outer regions of the
L/SMC and the Magellanic Bridge. The LMC star clusters span a narrower range of distances than
those in the SMC, suggesting that the SMC is more elongated than the LMC along the LOS.
This spatial pattern traced by star clusters is also seen from other galaxy components which show
that the Magellanic Bridge, with its onset in the SMC \citep{petal15a}, connects both Clouds \citep{wagnerkaiser17}, and that the SMC is elongated along the LOS \citep{jacyszynetal2017,nideveretal2019,massanaetal2020}. We point out that the novel picture 
of SMC star clusters spanning  a range of $\sim$ 15 kpc in distance along the LOS would not 
have been disentangled if a mean distance modulus had been adopted while analyzing the star 
cluster CMDs.

\begin{figure}
\includegraphics[width=\columnwidth]{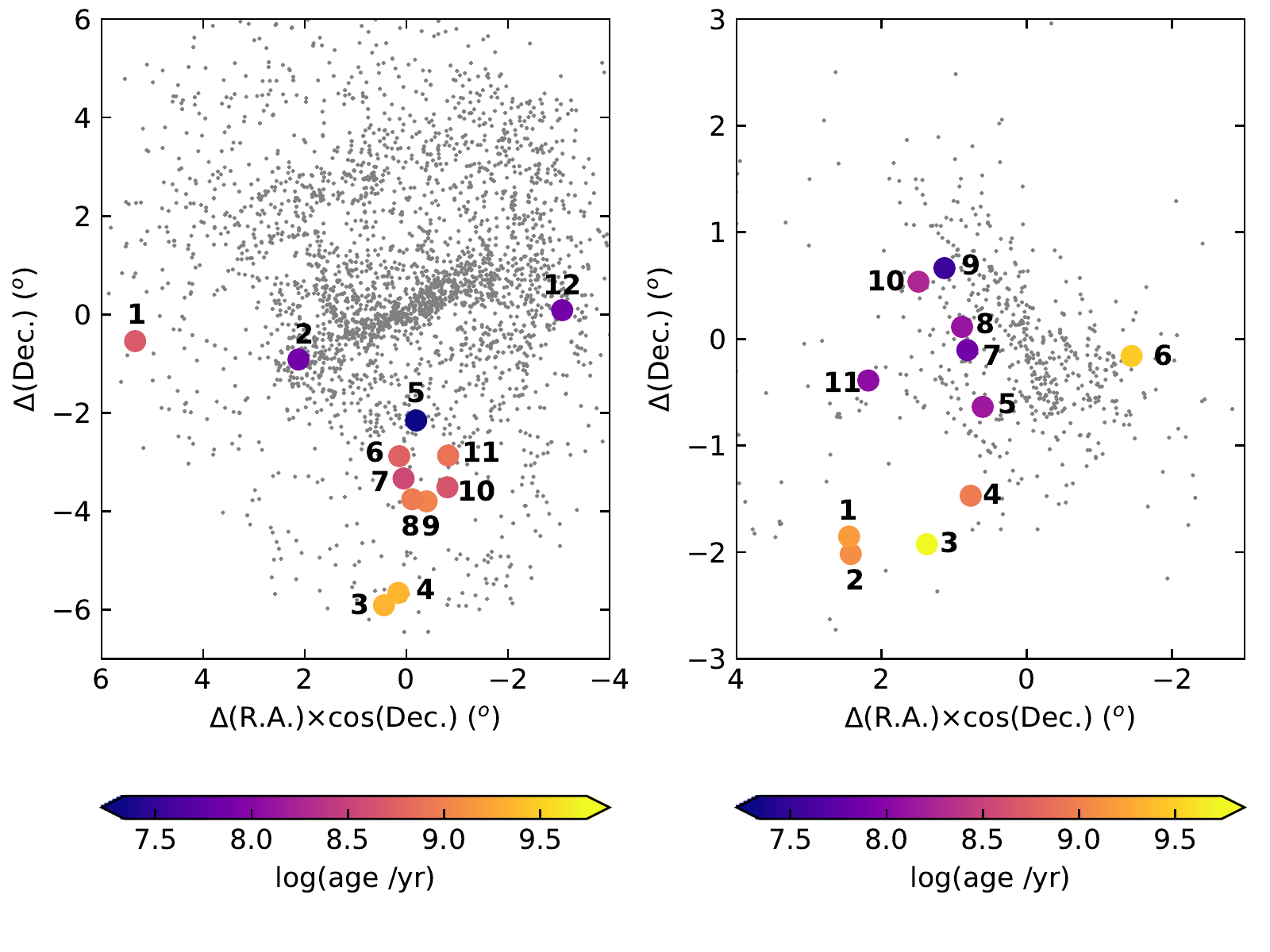}
\caption{Same as Figure~\ref{fig:fig5}, color-coded according to the star cluster ages.}
\label{fig:fig6}
\end{figure}

The age estimates of these star clusters (see Figure~\ref{fig:fig6}) are also worthy of
discussion. In general terms they confirm the outside-in formation scenario \citep{cetal08,meschin14},
in which more metal-poor star clusters formed first and the youngest ones were born from the gas
that collapsed to the innermost regions. Old star clusters formed at the core of the galaxy
had more chances to be disrupted. Hence, a spatial age distribution similar to an age gradient is 
observed.  There is some exceptions to this simple view, that arise as a consequence of the galaxy interactions. Relatively young star clusters with a metallicity content of those formed in the LMC
bar or inner disk were found in the outer LMC disk, where older and more metal-poor ones are 
expected to survive \citep{p16}. \citet{piattietal2018a} showed that such objects could have been 
born in the innermost LMC regions and then scattered to the outer LMC disk. Ram pressure 
interaction between the LMC and the Milky Way and between both Magellanic Clouds could also 
triggered star cluster formation in the periphery of the Clouds \citep{siteketal2016,piattietal2018b}, 
while some old globular clusters could be associated to the accreted satellite populations of the 
LMC \citep{martinetal2016,cernyetal2020}.

Within our studied star cluster sample, Field\,4-01 is located in the so-called West halo of
the SMC \citep{detal14}, a region placed on the opposite side of the bridge that was predicted 
by \citet{db2012} models and most likely has a tidal origin linked to the dynamical history of the 
Magellanic Clouds. Most of the known star clusters grouped in this area are in general of 
intermediate-age or older \citep{diasetal2016}, so that Field\,4-01 can be now added to this
group (log(age /yr)=9.45). Below the onset of the Magellanic bridge and somehow 
superimposed to it (to the southeast from the SMC center), there is also a region of moderately 
old star clusters \citep{petal07,p11c,p11b}. We estimated an age of 5.5 Gyr (log(age /yr)= 9.74)
for Field\,12-01 that matches very well the ages of this group of star clusters. Other two star
clusters, Field\,16-01 and 16-02, also located in this region, are a bit younger though
 (log(age /yr)$\sim$ 9.2). We note that Field\,12-01 is located at the LMC distance, so that it could be
a halo SMC star cluster affected by the LMC gravitational field \citep{carpinteroetal2013}.
By  using the elliptical framework devised by \citet{petal07d} to reflect more meaningfully the 
flattening of the galaxy, the remaining studied SMC star clusters are located to the east of the
SMC center at semi-major axes of $\sim$ 3$\degr$ and beyond.  They resulted to be young 
objects (age $\la$ 200 Myr), with only one exception, Field\,11-03  (log(age /yr) $\sim$ 9.0).  While the
older one is in agreement with the average age of star clusters in that region
\citep[ log(age /yr) $\ga$ 9.0 ][]{diasetal2016}, the younger one would seem more tightly related to 
the onset of the Magellanic bridge \citep{petal15a}. As for the spatial distribution of ages of
LMC star clusters, that of Field\,55-01 (the easternmost star cluster in the studied sample) 
would seem to be that of a likely runaway object \citep{piattietal2018a}, while the other
star clusters agree with the presence of an age gradient; those farther from the LMC
center being older.

The chemical enrichment of the Magellanic Clouds has long been studied from theoretical and
observational approaches. Some of the most recurrent age-metallicity relationships used in this
field are the models computed by \citet[][PT98]{pt1998}, which predict  intensive star formation and chemical enrichment during the inicial formation epoch which brought the metallicity up about -0.7 
dex and -1.3 dex, for the LMC and SMC, respectively. This turbulent period was subsequently 
followed by relative quiescence period to finally be disturbed by rapid burst of chemical enrichment 
about 3 Gyr ago  (log(age /yr) $\approx$ 9.5),  which brought the global metallicities up to their current values. Because of the
coincidence of the ages of L/SMC star clusters formed at that time, it has been argued that the bursting 
formation events were caused by the interaction between both Clouds \citep{p11a,p11b,p12a}. 
\citet[][BT12]{bt12}  also presented a bursting model for the LMC with some different ingredients. 
Closed-box model  of chemical evolution presented by 
\citet[][, G98]{getal98} and \citet[][, closed-box]{dh98}  predicted  gradual increase of star 
formation and metal abundances over time. The major merger scenario for the SMC was 
proposed by \citet[][TB09]{tb2009}. The model predicts that major merger occurred $\approx$7.5 Gyr 
ago and was calculated for three cases: no merger - (TB09-01), one-to-one merger (TB09-02), 
and one-to-four merger (TB09-03). From an observational point of view, \citet[][HZ09]{hz09} and
\citet[][HZ04]{hz04} built age-metallicity relationships for the LMC and SMC, respectively based 
on $UBVI$ photometry from Magellanic Clouds Photometric Survey, while \citet[][PG13]{pg13}
constructed the age-metallicity relationships for L/SMC star clusters using Washington photometry.

\begin{figure}
\includegraphics[width=\columnwidth]{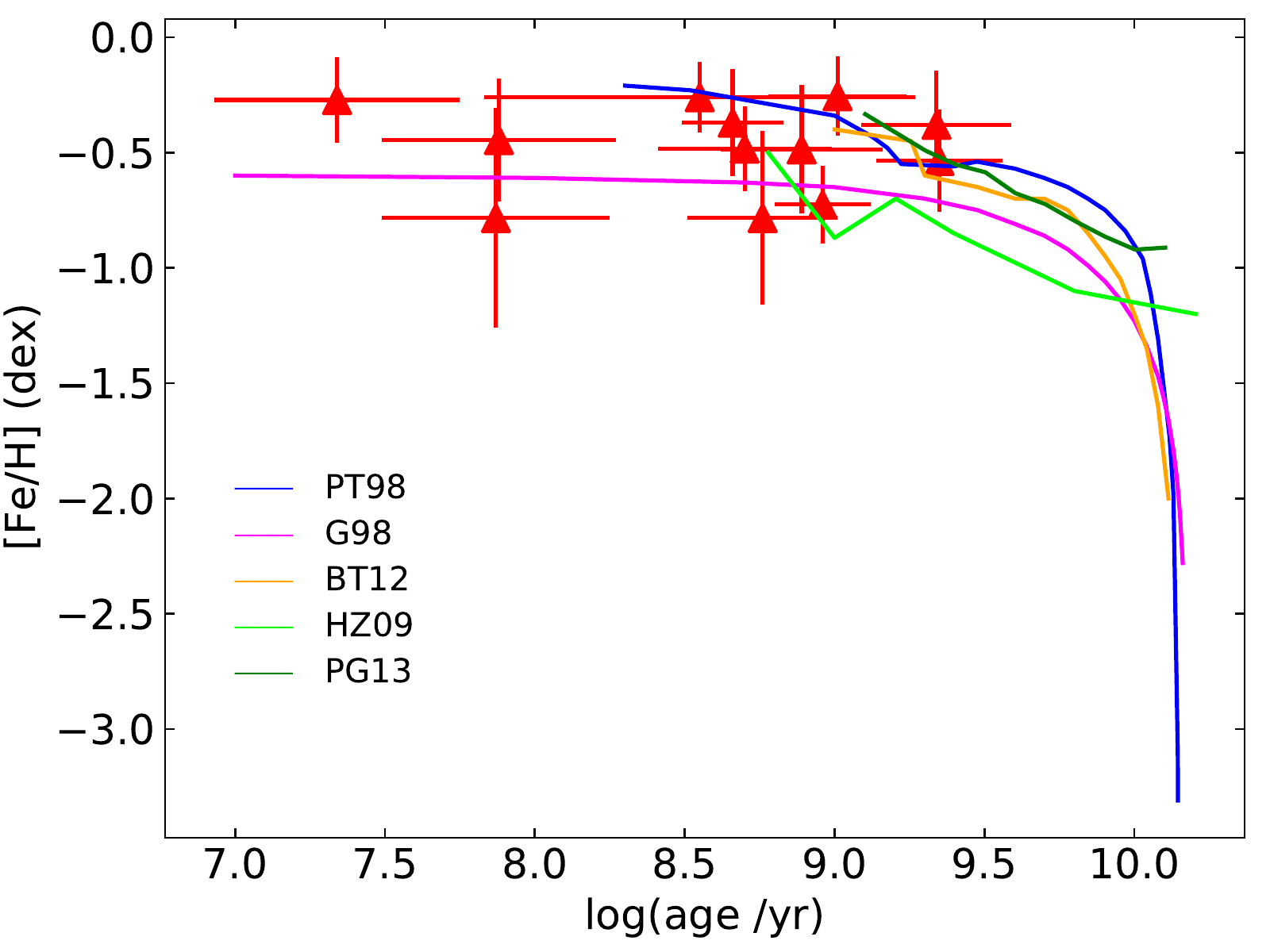}
\includegraphics[width=\columnwidth]{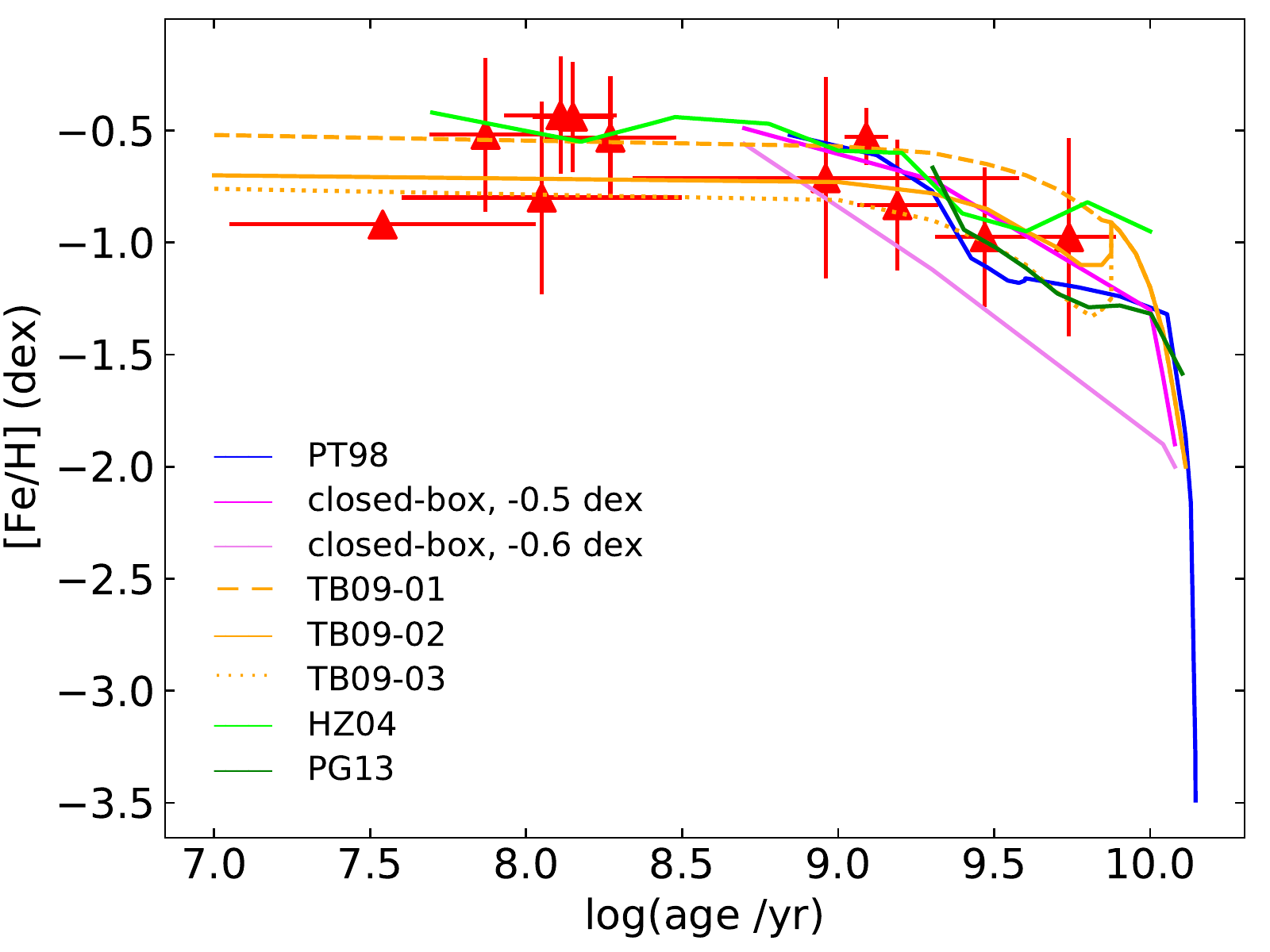}
\caption{Age-metallicity relationships for the LMC (top panel) and SMC (bottom panel). Triangles
represent star clusters studied in this work (error bars included, see Table~\ref{tab:tab1}), while 
the colored lines correspond to different theoretical or observed age-metallicity relationships: 
\citet[][PT98]{pt1998}; \citet[][G98]{getal98}; \citet[][BT12]{bt12}; \citet[][HZ09]{hz09}; 
\citet[][PG13]{pg13}; \citet[][closed-box]{dh98}; \citet[][TB09]{tb2009}; and \citet[][HZ04]{hz04}.
}
\label{fig:fig7}
\end{figure}

Figure~\ref{fig:fig7} shows the above listed age-metallicity relationships for the Magellanic Clouds
with the present studied star clusters superimposed. They span a quite wide range of ages, from
very young star clusters (log(age /yr) $\sim$ 7.5) up to intermediate-age ones (log(age /yr) $\sim$
9.5). In the LMC, none of the discovered star clusters turned out to be older than $\sim$ 2.5 
Gyr (log(age /yr) $\approx$ 9.4),
which is in agreement with the general consensus of the existence of a star cluster age gap in the 
LMC, from $\sim$ 4 Gyr  (log(age /yr) $\approx$ 9.6) up to the oldest globular clusters' ages 
\citep{oetal91,richetal2001,piattietal2009,pg13}. The metallicities of the L/SMC star clusters
are within the theoretically predicted and observed boundaries. Perhaps, the most noticeable
feature is the existence of young star clusters (age $\la$ 100 Myr) with relatively low metal
content in both Magellanic Clouds ([Fe/H] $\sim$ -0.7 dex)), similar to $\sim$ 1 Gyr old 
 (log(age /yr) $\approx$ 9.0)
metal-poor star clusters. This might imply that a relatively metal deficient gaseous flow could 
have existed between both Clouds during the last Gyrs,  also responsible of the Magellanic Stream 
and Leading Arm, that triggered star cluster formation \citep{ruizlaraetal2020,tsugeetal2020}. 
On the other
hand, chemical enhancement has reached slightly different metallicity levels. The iron-to-hidrogen
ratios increased on average up to -0.30$\pm$0.15 dex nad -0.50$\pm$0.20 dex, in the LMC and 
SMC, respectively. The present results illustrate that the Magellanic Clouds are more complex
galactic systems than previously known. The consideration of initial galaxy formation and later 
interactions between them and with the Milky Way guide us towards a better understanding of 
their present age and metallicity distributions of star clusters throughout the entire Magellanic
system.

\section{Concluding remarks}

We conducted analyses to obtain fundamental parameters estimates of 24 Magellanic Cloud star
cluster candidates recently discovered by \citet{p17a} using the SMASH DR2 database. We
find that all candidates resulted to be genuine physical systems, with the sole exception of
one candidate called Field\,30-02. We arrive to such a conclusion once the observed
star cluster CMDs were carefully decontaminated from field stars, and the cleaned CMDs
were compared to synthetic CMDs generated for thousand combinations of ages, distances,
metallicities, star cluster mass and binary fractions. The parameter of the best-matched synthetic
CMDs obtained from a likelihood approach were adopted as the star cluster astrophysical 
properties. In doing the comparison between observed and synthetic CMDs, we used only 
stars that passed the cleaning procedure and were assigned membership probabilities higher 
than 50$\%$. The use of a parallel tempering Bayesian MCMC algorithm to explore the
multi-parameter space allowed us to avoid typical constraints of adopting mean Magellanic
Cloud distances and metallicities in studies of star clusters. Those assumptions provided with
a limited picture of the Magellanic Clouds, where extended halos and tidally distorted peripheries 
are not distinguished. 

Indeed, the present star cluster sample spans a wide range of distances, from those
star clusters located behind the SMC, going through those placed in between both Clouds along
the Magellanic bridge, to those in from of the LMC. Such spatial distribution is by itself a
witness of interaction between both Clouds. Their estimated ages also tell us about a mixture
of formation episodes. Some clusters were born according to the outside-in formation 
scenario, where older star clusters are more commonly seen in outer galaxy regions
\citep{gallartetal2008,carreraetal2011}. However,
because of the interaction between both Magellanic Clouds and that of the Clouds with the
Milky Way, gas flows could have existed, initially feeding the outer regions where young
star cluster formed out of them. We find evidence of such formation phenomenon from the
identification of star clusters with different ages populating the same galaxy region and
star clusters that are found projected toward regions with associated stellar ages and
metallicities different from those of the star clusters. The estimated ages and metallicities 
confirm the general accepted evolution of the chemical enrichment in the Magellanic Clouds.
We find from the estimated star cluster metallicities another indicator of the existence of gaseous 
flows between  these galaxies. There exist in both Clouds very young clusters ($\sim$ 30 Myr),
located in their outer regions, with metal abundances as metal deficient as the most 
metal-poor star clusters with ages  $\la$ 1 Gyr  (log(age /yr) $\approx$ 9.0). The most metal-rich young star clusters have
slightly different [Fe/H] values, those of the LMC being more metal-rich.

Comparing the present-day masses of the studied star clusters with those of Milky Way open 
clusters with similar ages located in the solar neighborhood 
\citep[distance to the Sun $<$ 1.8 kpc,][]{joshietal2016}, we find that the studied
Magellanic Cloud  clusters are in general similar or
more massive than open clusters. 
%This likely tells us that mass disruption by tidal effects
%has been more important in the Milky Way \citep{pm2018,piattietal2019b}, 
%even though some of the studied Magellanic Cloud star clusters were subject of the
%interactions between both galaxies. 
Their binary frequencies is on average $q$=0.55,
independent of the star cluster mass.

\begin{acknowledgements}
I thank the referee for the thorough reading of the manuscript and
timely suggestions to improve it. 
This research uses services or data provided by the Astro Data Lab at NSF's National Optical-Infrared Astronomy Research Laboratory. NSF's OIR Lab is operated by the Association of Universities for 
Research in Astronomy (AURA), Inc. under a cooperative agreement with the National Science 
Foundation.
\end{acknowledgements}

%\bibliographystyle{aa}
%\bibliography{paper} % if your bibtex file is called paper.bib

%\input{paper.bbl}

\end{document}